\documentclass[prb,
superscriptaddress,showpacs,amsmath,amssymb]{revtex4}

\begin{document}

\author{G.E.~Volovik}
\affiliation{Low Temperature Laboratory, Aalto University,  P.O. Box 15100, FI-00076 Aalto, Finland}
\affiliation{Landau Institute for Theoretical Physics, acad. Semyonov av., 1a, 142432,
Chernogolovka, Russia}

\title{Analog Sommerfeld law in quantum vacuum}

\date{\today}

\begin{abstract}
The activation temperature $T$ in the de Sitter environment is twice larger than the Gibbons-Hawking temperature, related to the cosmological horizon. We consider the activation temperature as the local temperature of the de Sitter vacuum, and construct the local thermodynamics of the de Sitter state. This thermodynamics includes also the gravitational coupling $K$ and the scalar Riemann curvature ${\cal R}$ as the thermodynamically conjugate variables. These variables modify the thermodynamics of the Gibbs-Duhem relation in the de Sitter state. The free energy density is proportional to $-T^2$, which is similar to that in the non-relativistic  Fermi liquids and in relativistic matter with equation of state $w=1$.  The local entropy is proportional to the local temperature, while the total entropy inside the cosmological horizon is  $A/4G$, where $A$ is the area of the horizon. This entropy is usually interpreted as the entropy of the cosmological horizon. We also consider the possible application of the de Sitter thermodynamics to the Schwarzschild-de Sitter black hole and to black and white holes with the de Sitter cores.
\end{abstract}

\maketitle

\tableofcontents

\section{Introduction}

The vacuum of the de Sitter spacetime is characterized  by the local temperature  $T=H/\pi$, where $H$ is the Hubble parameter, see Ref. \cite{Volovik2023} and references therein. This temperature describes the thermal processes of decay of the composite particles and the other activation processes, which are energetically forbidden in the Minkowski spacetime, but are allowed in the de Sitter background, see also Refs. \cite{Maldacena2015,Reece2023,Maxfield2022}. In particular, this temperature determines the probability of the ionization of an atom in the de Sitter environment, $\exp(-E/T)$, where $E$ is the ionization potential. This activation temperature is twice the Gibbons-Hawking\cite{GH1977} temperature $T_{\rm GH}=H/2\pi$ of the cosmological horizon, $T=2T_{\rm GH}$. 
As distinct from the $T_{\rm GH}$, the activation temperature has no relation to the cosmological horizon. It describes the local processes, which take place far away from the horizon. 

However, it is not excluded that the processes related to event horizon may also contain the factor 2 due to entanglement between the observed particle and its partner hidden behind the horizon.\cite{tHooft2023,tHooft2022,tHooft2017,tHooft1984} The similar entanglement may produce the factor 2 for the processes related to the de Sitter cosmological horizon.\cite{Volovik2022a} This factor also appears in condensed matter analogs, where both the created (quasi)particle and its partner behind the acoustic horizon are observed by external observer.

If $T=H/\pi$ is the local temperature in de Sitter spacetime, the natural question is: does it determine the local thermodynamics of the de Sitter vacuum? In this paper we discuss this thermodynamics, including the Gibbs-Duhem relation of the de Sitter state, its free energy, effective pressure and local entropy. This can be extended to quasi-equilibrium states: vacuum+matter. Such states can be characterized by two temperatures: the temperature of the vacuum component and the temperature of the matter degrees of freedom. The main difference between thermodynamics of the two components of the system is that the vacuum component contains extra thermodynamic variables describing the gravitational degrees of freedom. 
These are the gravitational coupling $K=1/16\pi G$ (here we use units $c=\hbar=1$) and the scalar Riemann curvature ${\cal R}$, which are the thermodynamically conjugate variables. These variables enter the modified Gibbs-Duhem relation.

In Sec \ref{Gravastars} we apply the de Sitter thermodynamics to the gravastar -- the black hole with the de Sitter core, where the thermodynamics of black hole is modified by the thermodynamics of the de Sitter vacuum in its core. We show how the local entropy of the de Sitter vacuum cancels the entropy of the black hole horizon, and consider the similar cancellation for the "anti-gravastar" -- the white-hole with the de Sitter core.

\section{Thermodynamics of the de Sitter state}

 In the Painlev\'e-Gullstrand (PG) form\cite{Painleve,Gullstrand}  the metric in the de Sitter expansion is
 \begin{equation}
ds^2= - dt^2 +   (dr - v(r) dt)^2+r^2 d\Omega^2\,,
\label{PG}
\end{equation}
where the shift velocity is $v(r)=Hr$, and $H$ is the Hubble parameter.  

\subsection{Local temperature and local entropy}

From Friedmann equations of general relativity it follows that the vacuum energy density (which is the cosmological constant $\Lambda$) expressed in terms of the activation temperature $T=H/\pi$ is:
\begin{equation}
 \epsilon_{\rm vac}=\Lambda=\frac{3}{8\pi G}H^2=\frac{3\pi}{8G}T^2\,.
\label{dSEnergyDensity}
\end{equation}

We assume, that as distinct from the Gibbons-Hawking temperature related to the cosmological horizon, the temperature  $T=H/\pi$ is the local temperature of the de Sitter vacuum. Then we can determine  the free energy density of the de Sitter vacuum, $F=\epsilon_{\rm vac} - T d\epsilon_{\rm vac}/dT$, and thus the entropy density $s_{\rm vac}$ in the de Sitter vacuum:
\begin{equation}
s_{\rm vac}= - \frac{\partial F}{\partial T} =\frac{3\pi}{4G}T=\frac{3}{4G}H\,.
\label{dSEntropyDensity}
\end{equation}
Here we used the quadratic dependence of vacuum energy density on temperature in Eq.(\ref{dSEnergyDensity}). 

\subsection{Modified Gibbs-Duhem relation}

The quadratic dependence of vacuum energy on temperature is also important for consideration of the thermodynamic Gibbs-Duhem relation for quantum vacuum. It leads to the reformulation of the vacuum pressure.
The conventional vacuum pressure $P_{\rm vac}$ obeys the equation of state $w=-1$ and  enters the energy momentum tensor of the vacuum medium in the form:
\begin{equation}
T^{\mu\nu}= \Lambda g^{\mu\nu} = {\rm diag}( \epsilon_{\rm vac},  P_{\rm vac},   P_{\rm vac}, P_{\rm vac}) \,\,, \,\,
P_{\rm vac}=-\epsilon_{\rm vac}\,.
\label{EnergyMomentum}
\end{equation}
In the de Sitter state the vacuum pressure  is negative, $P_{\rm vac}=-\epsilon_{\rm vac}<0$. 

This pressure $P_{\rm vac}$ does not satisfy the thermodynamic Gibbs-Duhem relation, $Ts_{\rm vac}=  \epsilon_{\rm vac}+ P_{\rm vac}$, because the right hand side of this equation is zero.
The reason for that is that in this equation we did not take into account the gravitational degrees of freedom. Earlier it was shown, that gravity contributes to thermodynamics with the pair of the  thermodynamically conjugate variables:  the gravitational coupling $K=\frac{1}{16\pi G}$ and the Riemann curvature  ${\cal R}$, see Refs.\cite{KlinkhamerVolovik2008c,Volovik2022,Volovik2020}. This is because the Einstein-Hilbert action contains the gravitational term  $K{\cal R}$, and its contribution to thermodynamics is somewhat similar to the work density.\cite{Hayward1998,Hayward1999,Jacobson1995,Odintsov2023a} 
 The gravitational thermodynamic variables allow us to write the modified  Gibbs-Duhem relation:
 \begin{equation}
Ts_{\rm vac}=  \epsilon_{\rm vac}+ P_{\rm vac} -K{\cal R}\,.
\label{GibbsDuhem}
\end{equation}
This equation is obeyed, since $\epsilon_{\rm vac}+ P_{\rm vac}=0$ and ${\cal R}=-12H^2$.

Eq.(\ref{GibbsDuhem}) suggests that one may introduce the effective pressure, which is modified by gravitational degrees of freedom:
\begin{equation}
P= P_{\rm vac} -K{\cal R} \,.
\label{EffectiveP}
\end{equation}
Then the conventional  Gibbs-Duhem relation   is satisfied:
\begin{equation}
Ts_{\rm vac}=  \epsilon_{\rm vac}+ P\,.
\label{EffectiveGibbs}
\end{equation}

The effective de Sitter pressure $P$ is positive, $P=\epsilon_{\rm vac}>0$, and satisfies equation  of state $w=1$, which is similar to matter with the same equation of state. As a result, due to the gravitational degrees of freedom, the de Sitter state has many common properties with the non-relativistic Fermi liquid, where the thermal energy is proportional to $T^2$, and with the relativistic matter with $w=1$. This means that in thermodynamics the de Sitter vacuum behaves as the stiff matter introduced by Zel'dovich,\cite{Zeldovich1962} where the speed of sound is equal to the speed of light, $s^2=c^2 dP/d\epsilon_{\rm vac}=c^2$. 

\subsection{Entropy of cosmological horizon}

In this vacuum thermodynamics, the total entropy in the volume $V_H$ surrounded by the cosmological horizon with radius $R=1/H$ is
\begin{equation}
s_{\rm vac}V_H=\frac{4\pi R^3}{3} s _{\rm vac}= \frac{\pi}{GH^2}=\frac{A}{4G} \,,
\label{dSEntropy}
\end{equation}
where $A$ is the horizon area. This corresponds to the Gibbons-Hawking entropy of  the cosmological horizon. However, here it is the thermodynamic entropy coming from the local entropy of the de Sitter quantum vacuum, rather than the entropy of the horizon degrees of freedom. The vacuum thermodynamics suggests some  holographic connection between these two entropies.

\subsection{Entropy of black hole horizon}

The modified Gibbs-Duhem relation in Eq.(\ref{GibbsDuhem}) is applicable also to the thermodynamics of black holes.
As distinct from the de Sitter state, the black hole is the compact object, and its thermodynamics connects the global parameters, such as mass $M$, entropy of horizon, total electric charge $Q$ and total angular momentum $J$. This global thermodynamics can be described by the integral form of the  Gibbs-Duhem relation in Eq.(\ref{GibbsDuhem}). The curvature here comes from the central singularity:\cite{Balasin1993} 
\begin{equation}
{\cal R}=8\pi MG\,\delta({\bf r})\,.
\label{CentralSingularity}
\end{equation}
Since the energy density here is $\epsilon=M\delta({\bf r})$,   the integration of the right-hand-side of Eq.(\ref{GibbsDuhem}) over space gives for the Schwarzschild black hole: 
\begin{equation}
T_{\rm BH} S_{\rm BH} =M-\int d^3 r \sqrt{-g} K{\cal R}= \frac{M}{2}\,.
\label{BlackGibbs}
\end{equation}
Here $T_{\rm BH}=\frac{1}{8\pi MG}$ is the Hawking temperature and $S_{\rm BH}=A/4G$ is the Bekenstein-Hawking entropy. The Eq.(\ref{BlackGibbs}) is valid also for the white hole, where temperature and entropy are opposite to that of the black hole with the same mass,\cite{Volovik2022} 
 $T_{\rm WH}(M)=- T_{\rm BH}(M)$ and  $S_{\rm WH}(M)=- S_{\rm BH}(M)$.

\subsection{Entropy of the Schwarzschild-de Sitter cosmological horizon}

Let us consider the possible application of the modified Gibbs-Duhem relation to the Schwarzschild-de Sitter (SdS) black hole. We discuss the simple case of the Nariai limit, when the black hole horizon approaches the cosmological horizon, $r_b \rightarrow r_0-0$ and  $r_c \rightarrow r_0 + 0$, where
\begin{equation}
r_0=(GM/H^2)^{1/3} =\frac{1}{\sqrt{3}H} \,.
\label{r0}
\end{equation}
 In this limit the temperatures of the horizons approach the Bousso-Hawking value:\cite{BoussoHawking1996} 
\begin{equation}
T_b=T_c=\frac{\sqrt{3}}{2\pi}H=\frac{1}{6\pi GM}\,.
\label{GlobalT}
\end{equation}

 The entropy of the cosmological horizon $S_{\rm c} $ can be obtained by integration of the right-hand-side of Eq.(\ref{GibbsDuhem}) over space inside the cosmological horizon with radius $r_0$:
\begin{equation}
T_{\rm c} S_{\rm c} =M-\int _{r<r_0}d^3 r \sqrt{-g} K{\cal R}  \,.
\label{SdSGibbs}
\end{equation}
In the Nariai limit this gives:
\begin{equation}
T_{\rm c} S_{\rm c} =M-\int_{r<r_0} d^3 r \sqrt{-g} K{\cal R}  = M- \frac{M}{2} + \frac{3 H^2}{4\pi} \, \frac{4\pi r_0^3}{3}= \frac{3}{2}M\,,
\label{SdSGibbs2}
\end{equation}
and one obtains the entropy of the cosmological horizon, which agrees with the Gibbons-Hawking entropy:
\begin{equation}
S_{\rm c}=\frac{3M}{2T_{\rm c}} =\frac{\pi r_0^2}{G}= \frac{A}{4G}\,.
\label{SdSentropy}
\end{equation}

\subsection{Sommerfeld law}

Since the thermodynamics of the de Sitter state with the thermal energy $\epsilon_{\rm vac}\propto T^2$ is similar to the thermodynamics of the Fermi liquid, let us try to exploit this connection. One of the directions is the Sommerfeld law in Fermi liquid, which states that the entropy for one atom of the Fermi liquid is $S\propto T/E_F$, where $E_F$ is Fermi energy. We do not know what are  the "atoms of the vacuum", but from Eq.(\ref{dSEntropyDensity}) it follows that the entropy density of the vacuum  
$s_{\rm vac} \sim T/l_{\rm P}^2 \sim (T/E_{\rm P})/l_{\rm P}^3$, where $l_{\rm P}=\sqrt{G}$ is the Planck length and $E_{\rm P}$ is the Planck energy.
This suggests that the density of the "atoms of the vacuum" is $n_{\rm P} \sim 1/l_{\rm P}^3$ and entropy per  "atom of the vacuum" is:
\begin{equation}
S= \frac{s_{\rm vac}}{n_{\rm P}} \sim s_{\rm vac} l_{\rm P}^3 \sim \frac{T}{E_{\rm P}}  \,.
\label{Sommerfeld}
\end{equation}
  Eq.(\ref{Sommerfeld}) is the full analog of the Sommerfeld law for Fermi liquid. This analogy also suggests that the corresponding density of states in the quantum vacuum (the analog of density of states at the Fermi level $N_F\sim mp_F$  in Fermi liquids) is $N_P \sim E_{\rm P}^2$. For bosonic and fermionic degrees of freedom of this quantum vacuum the density of states is $N_P =9/4\pi G$ and $N_P =9/2\pi G$ correspondingly, which can be compared with the value  $N_P =3\pi/G$ suggested in Ref. \cite{Chu2023}. This huge density of states leads to a very large entropy of the de Sitter state even for very small temperature of the vacuum.
 
 So,  the quantum vacuum looks as some specific form of the relativistic Fermi liquid. In particular,  the energy density of such "Fermi liquid" is zero under the following conditions:\cite{KlinkhamerVolovik2008a,KlinkhamerVolovik2008b} there is the full equilibrium at $T=0$, there is no matter and no external pressure. Then one has Minkowski vacuum with $\epsilon_{\rm vac} (H=0, T=0)=0$. The non-zero  energy density of the vacuum and thus the cosmological constant in Einstein equations appear only if these conditions are violated. In case of de Sitter expansion the nonzero vacuum energy density is in  Eq.(\ref{dSEnergyDensity}).

\subsection{Expansion vs Planckian dissipation}
 
The temperature $T=H/\pi$ is the local temperature, which matter experiences as an activation temperature.\cite{Volovik2023} On the other hand, in the non-equilibrium Universe, matter may have its own temperature.\cite{Vergeles2023}
In the present non-equilibrium Universe the temperature of matter  is much larger then the temperature of the vacuum, $T_{M}\gg T=H/\pi$, while the entropy density of the vacuum is dominating. Under this condition the expansion causes the decay of the energy density of matter in the conventional way, as follows from Einstein equations:
\begin{equation}
\dot \rho_M +3(1+w)H\rho_M=0 \,.
\label{decay}
\end{equation}

For the relativistic gas with equation of state $w=1/3$, the average energy of particles is $E \sim T_M \sim \rho_M^{1/4}$, and one has;
\begin{equation}
\dot E = - HE\equiv  - \frac{1}{\tau_E} E \,.
\label{decay2}
\end{equation}
That is why the energy relaxation time of matter  $\tau_E$ in the de Sitter environment is:
\begin{equation}
\frac{1}{\tau_E}= H =\pi T \,.
\label{decay3}
\end{equation}
At first glance this looks as the Planckian dissipation,\cite{Zaanen2019} with $\tau_E$ playing the role of the inelastic relaxation time. However, Eq.(\ref{decay3}) relates the relaxation of matter and the temperature of the vacuum $T$.
If we consider the temperature of matter, we have
\begin{equation}
\frac{1}{\tau_E}=\pi T  \ll T_M\,.
\label{decay4}
\end{equation}
This means that the Planckian dissipation takes place only in the limit of low temperature of matter, when $T_M \sim T$. This  probably suggests that the temperature of matter cannot be smaller than the temperature of the vacuum.

\section{From black hole to gravastar and white hole}
\label{Gravastars}
\subsection{Gravastar with contracting de Sitter interior}

Let us apply this approach to such type of gravastars,\cite{Chapline2003,Mazur2023,Mottola2023} which contain the de Sitter state in the whole region inside the black hole horizon. Such gravastars are stationary and have no Hawking radiation. In the Painlev\'e-Gullstrand form the metric is given by Eq.(\ref{PG}) with the following shift velocity:
\begin{eqnarray}
v(r)= -\sqrt{\frac{r_h}{r}} \,\,,\,\, r >r_h \,,
\label{v1}
\\
v(r)=- \frac{r}{r_h} \,\,,\,\, r < r_h \,.
\label{v2}
\end{eqnarray}
Here $r_h=1/(2MG)$ is the horizon radius.

We assume that it is natural that  the shift  velocity $v(r)$ is continuous across the horizon, i.e. there is no jump in the shift velocity, and only the gradient of the shift velocity $dv/dr$ experiences jump at the horizon. That is why the shift velocity $v(r)$ is negative both outside and inside the horizon.  Since the shift velocity is negative in the de Sitter core, this means that the de Sitter spacetime in this gravastar is contracting, $v(r)=Hr=-r/r_h<0$, i.e. the Hubble parameter is negative, $H<0$. 

In the contracting de Sitter vacuum the local temperature and the  local entropy  are negative.\cite{Volovik2023} From Eq.(\ref{dSEntropyDensity}) it follows that $s=3H/4G=-3/(4Gr_h)<0$.
Then the entropy of the whole de Sitter region in  Eq.(\ref{dSEntropy}) is $sV_h=-\frac{A}{4G}$, where $A$ is the area of horizon. The de Sitter entropy fully compensates the black hole entropy of the horizon:
 \begin{equation}
 S_{\rm gravastar}=sV_h +S_{\rm BH}=-\frac{A}{4G}+\frac{A}{4G} =0\,.
\label{Gravastar}
\end{equation}

Eq.(\ref{Gravastar}) supports the statement that there is no entropy and no Hawking radiation in this type of gravastar.  Also it is important that the effective vacuum pressure $P$ in de Sitter core in Eq.(\ref{EffectiveP}) is positive. This matches the pressure induced by the surface tension of the thin shell separating de Sitter core from the outer region.

The absence of entropy in this type of gravastar is natural, since such gravastars are fully static, and can be described in static Schwarzschild coordinates. In this case the gravastar can be considered as the intermediate state between the black hole and the white hole.\cite{Volovik2022}  
While the white hole has negative entropy  $S_{\rm WH}=-S_{\rm BH}=-\frac{A}{4G}$,  the intermediate
state—the fully static hole—has zero entropy, $S_{\rm gravastar}=0$.   

\subsection{Antigravastar: white hole with expanding de Sitter interior}

The related type of gravastar is the "antigravastar" with positive shift function:
\begin{eqnarray}
v(r)= \sqrt{\frac{r_h}{r}} \,\,,\,\, r >r_h \,,
\label{v1anti}
\\
v(r)= \frac{r}{r_h} \,\,,\,\, r < r_h \,.
\label{v2anti}
\end{eqnarray}
It is the white hole with expanding de Sitter in the inner region, where $v(r)=Hr$ with $H>0$.
In this case the inner region has positive local entropy, which cancels the negative entropy of white hole. As a result the total entropy of the antigravastar is also zero:
 \begin{equation}
 S_{\rm antigravastar}=sV_h +S_{\rm WH}=\frac{A}{4G}-\frac{A}{4G} =0
\,.
\label{AntiGravastar}
\end{equation}
 
 \subsection{de Sitter-Schwarzschild black holes}

There are the other models of the de Sitter-Schwarzschild black hole,\cite{Dymnikova2002,Dymnikova2011} in which the region with de Sitter vacuum is formed from the central singularity of the black hole. This region has the  inner "cosmological" horizon, which grows and finally merges with the outer horizon forming the gravastar.    The inner horizon here is the white hole horizon, and thus has negative entropy. After merging with event horizon, the entropy of gravastar becomes zero. Bekenstein-Hawking entropy related to horizon is compensated by local entropy of the de Sitter vacuum. However. in this model the critical value of mass, at which two horizons merge, is of the Planck energy scale.

 The gravastar as the vacuum star can be considered using the special variable $q$, which describes the quantum vacuum.\cite{KlinkhamerVolovik2008a,KlinkhamerVolovik2008b}
 The $q$-field is in particular represented by the 4-form field. It was the 4-form field, which was originally used by Hawking for consideration of the problem related to vacuum energy and cosmological constant.\cite{Hawking1984} With the vacuum $q$-field it was shown, that while the gravastar is formed, it does not contain  the de Sitter core.\cite{Zubkov2023} Maybe this is because of the instability of the de Sitter state with negative temperature. On states with negative temperature see Ref.\cite{Volovik2021}  and references therein. 
 Also, the fate of the Sitter-Schwarzschild black hole depends of the type of the de Sitter core, which can be either expanding or contracting.

 \section{Conclusion}
 
 In conclusion, the local thermodynamics of de Sitter spacetime, with local temperature, local entropy, and local Gibbs-Duhem relation, can be constructed. For that one must use also the thermodynamics of the gravitational degrees of freedom: the gravitational coupling $K$ and Riemann scalar curvature ${\cal R}$ as thermodynamically conjugate variables. With these variables added, the thermodynamics becomes similar to the thermodynamics of the Fermi liquids with the entropy density in Eq.(\ref{dSEntropyDensity}), which is linear in the local temperature $T$. 
 
 The local de Sitter temperature is determined by the action of the expanding Universe on the matter degrees of freedom: it describes the processes of activation, such as the thermal process of the ionization of atoms in the de Sitter environment. This activation temperature has no relation to the cosmological horizon, and is twice larger than the Hawking temperature related to the horizon. Nevertheless, the total entropy in the region inside the cosmological horizon is exactly the horizon entropy $A/4G$. This is a kind of the bulk-boundary correspondence, which illustrates the mystery of the de Sitter horizon:  "It seems fair to say that although black hole entropy remains highly enigmatic to this day, the entropy of a cosmological horizon, such as the de Sitter horizon, is only more mysterious."\cite{Witten2023} Anyway, if the bulk-boundary correspondence does take place and the local entropy gives rise to the Gibbons-Hawking entropy, it is not excluded that the other types of the local entropy may lead to the generalized entropy, such as discussed in Refs.\cite{Odintsov2023,Odintsov2022,Odintsov2021}.

Since the quantum de Sitter vacuum has its  local temperature, then in the quasi-equilibrium states with matter the system can be characterized by two temperatures: the temperature of the vacuum component and the temperature of matter degrees of freedom.  The present temperature of the vacuum component is much smaller than the temperature of matter degrees of freedom. For example, compared with the temperature of Cosmic Microwave Background (CMB) radiation it is $T_{\rm vac}\sim 10^{-30}T_{\rm CMB}$. But the entropy of the vacuum highly exceeds the entropy of matter due to large density of states in the quantum vacuum, $s_{\rm vac} \sim 10^{30} s_{\rm CMB}$. 
In the inflationary epoch the  situation can be different.\cite{Vergeles2023} 

The de Sitter thermodynamics determines the thermodynamics of the  such objects as gravastars and "anti-gravastars" with the de Sitter cores. It is shown how the local entropy of the de Sitter state in the inner region annihilates the global Bekenstein-Hawking entropy of the horizon. As a result, these gravastar objects are static, have zero entropy and no Hawking radiation.

 {\bf Acknowledgements}.   
 I thank Gerard ’t Hooft, Sergei Odintsov and Valerio Faraoni for discussions.
 This work has been supported by Academy of Finland (grant 332964).

\end{document}